\documentclass[conference]{IEEEtran}
\IEEEoverridecommandlockouts
\usepackage{cite}
\usepackage{amsmath,amssymb,amsfonts}
\usepackage{algorithmic}
\usepackage{graphicx}
\usepackage{soul}
\usepackage{float}
\usepackage{hyperref}
\usepackage{textcomp}
\usepackage{xcolor}
\usepackage{tcolorbox}
\usepackage{todonotes}
\usepackage{pifont} 
\usepackage{setspace} 
\usepackage{fontenc} 
\usepackage{ulem} 
\def\BibTeX{{\rm B\kern-.05em{\sc i\kern-.025em b}\kern-.08em
    T\kern-.1667em\lower.7ex\hbox{E}\kern-.125emX}}

\begin{document}

\title{Reimagining Self-Adaptation in the Age of Large Language Models}

\author{\IEEEauthorblockN{Raghav Donakanti, Prakhar Jain, Shubham Kulkarni,  Karthik Vaidhyanathan}
\IEEEauthorblockA{\textit{Software Engineering Research Center},  \textit{IIIT Hyderabad}, India\\
raghav.donakanti@students.iiit.ac.in, prakhar.jain@research.iiit.ac.in, shubham.kulkarni@research.iiit.ac.in, \\karthik.vaidhyanathan@iiit.ac.in
}}

\maketitle

\begin{abstract}
Modern software systems are subjected to various types of uncertainties arising from context, environment, etc. To this end, self-adaptation techniques have been sought out as potential solutions. Although recent advances in self-adaptation through the use of ML techniques have demonstrated promising results, the capabilities are limited by constraints imposed by the ML techniques, such as the need for training samples, the ability to generalize, etc. 
Recent advancements in Generative AI (GenAI) open up new possibilities as it is trained on massive amounts of data, potentially enabling the interpretation of uncertainties
and synthesis of adaptation strategies. 
In this context, this paper presents a vision for using GenAI, particularly Large Language Models (LLMs), to enhance the effectiveness and efficiency of architectural adaptation.
Drawing parallels with human operators, 
we propose that LLMs can autonomously generate similar, context-sensitive adaptation strategies through its advanced natural language processing capabilities. This method allows software systems to understand their operational state and implement adaptations that align with their architectural requirements and environmental changes. 
By integrating LLMs into the self-adaptive system architecture, we facilitate nuanced decision-making that mirrors human-like adaptive reasoning. A case study with the SWIM exemplar system provides promising results, indicating that LLMs can potentially handle different adaptation scenarios. 
Our findings suggest that GenAI has significant potential to improve software systems' dynamic adaptability and resilience. 
\end{abstract}


\section{Introduction}
\label{sec:intro}
Software systems face uncertainties, such as component
failures and variable user requests, affecting their Quality of
Service (QoS). Over the years, self-adaptation has emerged as
a potential solution to mitigate the uncertainties ~\cite{weyns2020introduction}. 

Off late, with advancements in computing and AI in general. Machine Learning (ML) has emerged as a key solution
for self-adaptation \cite{10.1145/3469440}, utilizing large datasets to identify
patterns and predict future states, thus enabling dynamic struc-
ture/behavioral adjustments. Despite its benefits, ML’s applica-
tion in self-adaptation requires extensive data and a thorough
understanding of ML principles. The rise of Generative AI
(GenAI), especially through Large Language Models (LLMs),
marks significant progress in this direction ~\cite{hou2023large}. LLMs enhance
machine understanding and human-like text generation capa-
bilities, showcasing their utility in diverse areas, especially for
complex decision-making given adequate context ~\cite{hou2023large, autogen}.

The concept of autonomic computing, as proposed by
Kephart and Chess \cite{mapek}, sought to enhance the autonomy of
software systems through various strategies. Despite these
efforts, a persistent challenge has been the ability of systems
to dynamically generate new configurations and components.
The advent of GenAI, particularly the capabilities of LLMs,
introduces the possibility of developing adaptation strategies
directly. This is supplemented by the fact that modern software
systems generate vast amounts of data, including logs, metrics,
and traces, which system administrators traditionally leverage
for tasks such as root cause analysis and resource allocation.
This data, encompassing code, APIs, JSON, XML, and more,
can be converted into various natural language formats, en-
abling Large Language Models (LLMs) to interpret the data
and generate adaptive responses akin to those formulated by
human experts. To this end, this paper introduces an innovative
framework, MSE-K (Monitor, Synthesize, Execute) that inte-
grates Large Language Models (LLMs) into the self-adaptation
process, enabling software systems to autonomously generate
and implement contextually relevant and actionable architec-
tural adaptation strategies aligned with their operational goals.
This approach seeks to overcome the limitations of current
self-adaptation mechanisms, paving the way for more efficient
and intelligent software systems \cite{SA2.0}.

As an initial validation, we performed evaluations of our
approach on SWIM \cite{SWIM}, an exemplar for simulating web
infrastructure through well-defined interfaces for monitoring
runtime metrics and executing adaptation strategies. Our initial
results indicate that incorporating LLMs into self-adaptive
systems has immense potential to improve the self-adaptability
of software systems and thereby guarantee performance in the
event of uncertainties.

\section{Related Works}
\label{sec:rel_works}
The integration of machine learning into self-adaptive systems has marked a significant advance in the field of software engineering. 
Gheibi et al. \cite{10.1145/3469440} have highlighted the use and key challenges of engineering ML-based adaptive systems. Approaches based on reinforcement learning for planning and adapting software systems were introduced in \cite{5069076} and \cite{10.1145/2701126.2701191}. \cite{9101287} combines ML for adaptation pattern selection and quantitative verification for decision feasibility. \cite{SA2.0} emphasizes the bidirectional relationship between self-adaptation and AI to manage unanticipated changes. \cite{lifelong_ml} proposes a novel approach to self-adaptation by integrating a lifelong ML layer into self-adaptive systems that use ML techniques, enabling dynamic tracking and updating of their learning models. This paper builds on existing research on integrating AI into self-adaptive systems by exploring the transformative potential of Generative AI, especially Large Language Models (LLMs), for managing uncertainties and devising adaptive strategies. It demonstrates LLMs' potential to improve decision-making in architectural adaptation.


\section{LLM-Based Architecture for Self-Adaptation}
\label{sec:llm}

\begin{figure}[t]
  \centering
  \includegraphics[width=\linewidth]
  {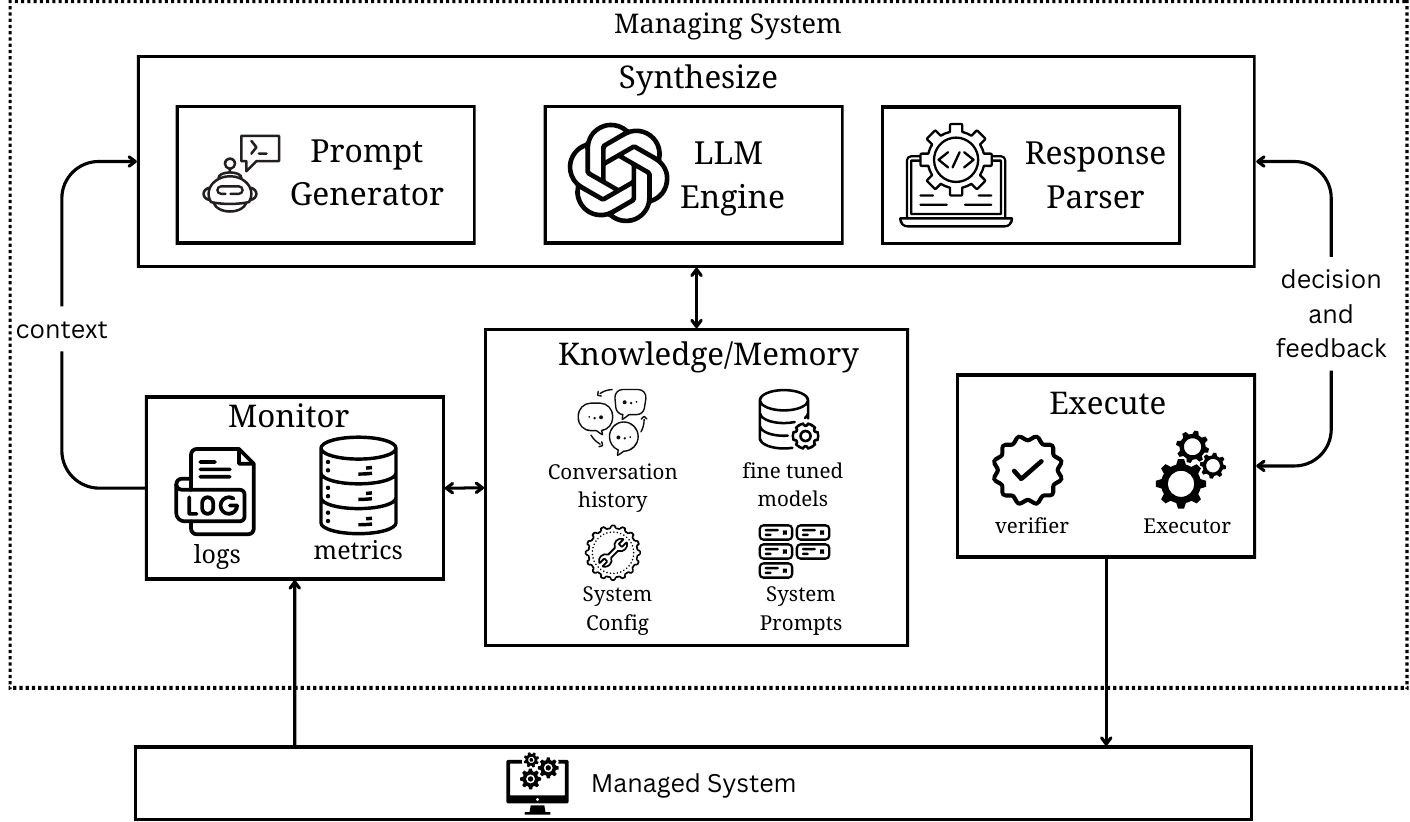}
  \caption{Approach}
  \label{fig:appr}
\end{figure}

The approach, as represented in Fig. \ref{fig:appr}, conforms to the conceptual model of a self-adaptive system. It 
consists of two subsystems, namely the \textit{Managed System} and \textit{Managing System}.
\textit{The Managed System} represents the operational software system equipped with probes for ongoing QoS monitoring and effectors for implementing adaptations.

\textit{The Managing System} is responsible for handling the adaptations. It consists of different components that implement the adaptation control loop, similar to MAPE-K~\cite{mapek}. However, in a traditional MAPE-K-based control loop of adaptation, the \textit{analyze} phase is responsible for constantly analyzing various aspects of the system to identify any possible deviations from the adaptation goals. Such deviation triggers the \textit{plan} phase, which generates adaptation decisions. In our approach, the analyze and plan phases are abstracted by the \textit{Synthesize} phase, which leverages LLM to interpret and generate adaptation decisions. This stems from the fact that modern-day LLMs have the ability to interpret information and further generate adaptation decisions. Hence, we envision an MSE-K loop as represented in Fig. \ref{fig:appr}. The rest of the section details the different components of the managing system.

\textbf{Monitor:} The monitor as in the case of the traditional MAPE-K approach is responsible for continuously monitoring the running software system. In our approach, the monitor activity is exploited to monitor and collect the system logs as well as the QoS metrics of the software system. To this end, it consists of two components \textit{logs collector}, which is responsible for collecting the system logs and a\textit{metric collector}, which collects different QoS metrics like response time, CPU utilization, etc. These data together form the context data, $C$, that represents the running state of the system at a given instant of time $t$. Further, these data are ingested into the Knowledge base for further processing and future use.

For instance, with respect to our case study on SWIM, the context data, $C$ is a set, such that $C = \{D_t, AS_{t}, U_t, R_t, AR_t, t\}$ where $D_t, \, AS_t, \, U_t, R_t \; and \; AR_t $ represent dimmer value, active servers, server utilization. average response time and arrival rate at time $t$ respectively.   

\textbf{Knowledge/Memory: }The Knowledge acts as a central storage for different types of knowledge required by different components of the managing system for performing the adaptations. It stores four types of information: \textit{i) Conversation History} which comprises tuples of past context data ($C$) and adaptation decision ($AD$) pairs. Hence the conversation history at timestep $t$, ${H_t} = [(C_1,AD_1),(C_2,AD_2)\ldots(C_t,AD_t)]$. The conversation history is updated with $C_t$ and $AD_t$ at every execution of the {\textit{Synthesize} step.}   
\textit{ii) Fine-tuned models} consist of a set of LLMS that can be used for different tasks by leveraging the notion of multi-agents~\cite {autogen}, where each LLM can be tasked with a role and responsibility. For instance, one LLM agent can be responsible for gathering insights from logs; another LLM agent can be responsible for interpreting and inferring information from system metrics and so on. These agents, as and when needed, can be leveraged by the \textit{LLM Engine} in the Synthesize component; \textit{iii) System Config} where the different configurations of the system are stored, such as the number of compute resources that are available, other hardware information, etc; \textit{iv) Prompts} Consists of different system prompts that provide the LLM with instructions on how to interpret the various data and further how to generate decisions along with the objective, $O$. They act as guidelines to ensure that the LLMs generate decisions that are relevant to the given context of a system and its objectives. Further, it also consists of a set of few-shot prompts which provide the LLM with a few examples of adaptation scenarios and the corresponding actions that need to be taken in such scenarios. These prompts further allow the LLM to learn how to generate an adaptation decision given a context. This also ensures that the LLM is more consistent when generating decisions.

\begin{figure}[t]
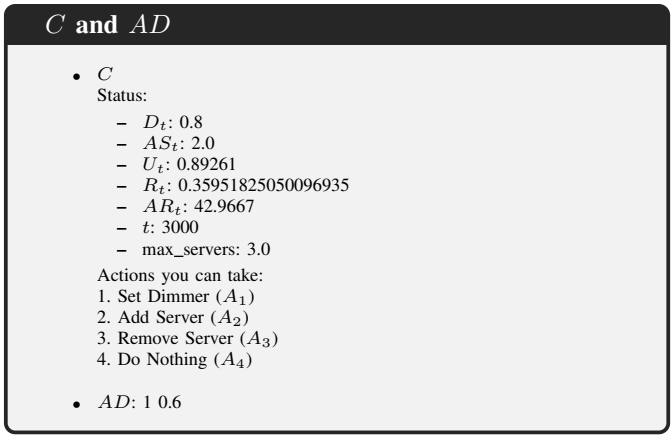

    \centering
    \begin{tcolorbox}[colback=black!5!white,colframe=black!85!white,title=\textbf{$C$ and $AD$}]
    \scriptsize 
    \begin{itemize}
        \item $C$ \\
        Status:
        \begin{itemize}
            \item $D_t$: 0.8
            \item $AS_t$: 2.0
            \item $U_t$: 0.89261
            \item $R_t$: 0.35951825050096935
            \item $AR_t$: 42.9667
            \item $t$: 3000
            \item max\_servers: 3.0
        \end{itemize}
    
        Actions you can take:\\
        1. Set Dimmer ($A_1$) \\
        2. Add Server ($A_2$)\\
        3. Remove Server ($A_3$)\\
        4. Do Nothing ($A_4$)\\

        \item $AD$: 1 0.6
    \end{itemize}
    \end{tcolorbox}
    \caption{Example of $C$ and $AD$ in SWIM}
    \label{fig:context_hist}
\end{figure}

\begin{figure}[t]
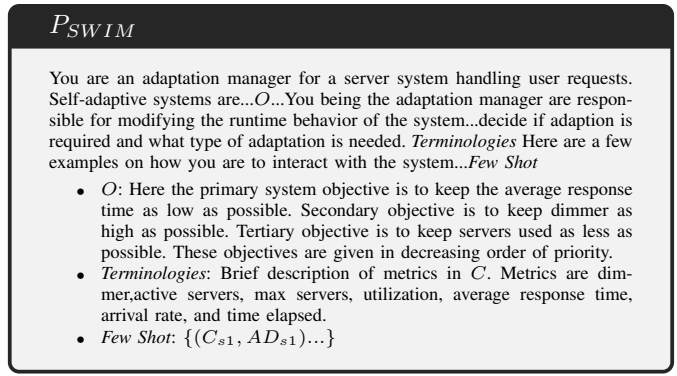

    \centering
    \begin{tcolorbox}[colback=black!5!white,colframe=black!85!white,title=\textbf{${P_{SWIM}}$} ]
       \scriptsize
    You are an adaptation manager for a server system handling user requests. Self-adaptive systems are...$O$...You being the adaptation manager are responsible for modifying the runtime behavior of the system...decide if adaption is required and what type of adaptation is needed. \textit{Terminologies} Here are a few examples on how you are to interact with the system...\textit{Few Shot}
    
    \begin{itemize}
        \item $O$: Here the primary system objective is to keep the average response time as low as possible. Secondary objective is to keep dimmer as high as possible. Tertiary objective is to keep servers used as less as possible. These objectives are given in decreasing order of priority.
        \item \textit{Terminologies}: Brief description of metrics in $C$. Metrics are dimmer,active servers, max servers, utilization, average response time, arrival rate, and time elapsed.
        \item \textit{Few Shot}: $\{(C_{s1},AD_{s1})...\}$
    \end{itemize}
    \end{tcolorbox}
    \caption{Example of $P_{SWIM}$}
    \label{fig:pswim}
\end{figure}

For instance, in the case of SWIM, we utilize \textit{Prompts} and \textit{Conversation History}. Our \textit{Prompts}, $P_{SWIM}$ (refer Figure \ref{fig:pswim}), consist of a description of what an adaptation manager is and what actions the LLM will be capable of performing, which contextualizes the LLM in our setting. This is paired with the objective the LLM needs to optimize for, followed by a brief description of various metrics in $C$ (refer Figure \ref{fig:context_hist}). Finally, a few-shot examples are provided.  

\textbf{Synthesize: } 
The Synthesize activity encompasses the Analyze and Plan phase of the traditional MAPE-K loop by interpreting the output of the \textit{Prompt Generator}, $P$ to generate the adaptation decision which is sent to the execute phase. It consists of three major components: 

\textit{i) Prompt Generator: } It is responsible for integrating the context, $C$ into the conversation history, $H$.  It then compiles $H$ and other long-term data stored in \textit{Knowledge} into a single prompt, $P$. Overall the prompt generator pulls together everything needed for the LLM to generate a decision, $P = \{C,H,\textit{long term data}\}$ 
Sometimes $P$ can lead to the LLMs context length being exceeded. This occurs because of how the transformer architecture can only process a finite number of tokens at once. To this, the \textit{Prompt Generator} may use strategies such as summarizing past decisions or truncating older context. It is then passed to the \textit{LLM Engine}.

\textit{ii) LLM Engine:} Understands the prompt $P$ passed into it and generates adaptation decisions. The conversation history, $H$ gives information pertaining to how previous $AD_t$ and $C_t$ variated through timesteps. The objective, $O$, helps in choosing adaptation decisions that build towards it. Using this data, the LLM will be able to generate an adaptation decision $AD$ (refer Figure \ref{fig:context_hist}) which is then pushed into $H$ and passed to \textit{Response Parser}. Each $AD$ may in turn consist of multiple actions. 

\textit{iii) Response Parser} Parses the $AD$ given and formats it. This formatting is necessary since the adaptation needs to comply with the communication format of the managed system. It also extracts arguments that the managed system may need from $AD$. 

For example, the $P$ that was provided at a timestep was of the form $P = \{C,H,P_{SWIM}\}$. The LLM generated response \textit{1 0.8}, which was further parsed by the \textit{Response Parser} to produce $1$ as the adaptation action and $0.8$ as the argument (dimmer value) for the action.

\textbf{Execute} is responsible for executing the adaptations on the running system. It further consists of two components. \textit{i) Verifier :} which verifies the given $AD$ by leveraging formal verification techniques to check quantitative properties on the system to determine if the proposed $AD$ aligns with our objective. If so the \textit{Executor} will execute the given $AD$ otherwise \textit{Synthesizer} will generate an alternative adaptation decision. \textit{{ii) Executor :}} Once the $AD$ is verified, the \textit{Executor} is responsible for the execution of the $AD$ within the managed system by modifying the system configuration or resources as specified by the $AD$.

For instance, if the \textit{LLM Engine} decides to add a server to handle higher traffic, the Verifier first confirms this action $A$ aligns with the objectives $O$. Upon verification, the Executor would allocate an additional server to accommodate the increased load. We have outlined the \textit{Executor} phase but haven't used the Verifier component in our approach for SWIM yet.



\section{Preliminary Results and Discussion}
\label{sec:results} 
The SWIM simulation was run in a dockerized environment, on a machine running a distribution of Ubuntu 22.04 LTS: AMD Ryzen 7 PRO 5850U at 4.5GHz and 16GB of RAM. The SWIM simulation was run with the world cup trace, starting with three servers and a dimmer value of 0.9, for the full length of the simulation (105 minutes). The adaptation manager was a python file, consisting of the interfaces to communicate with SWIM and the GPT-4 API (LLM). It interacts with SWIM through a TCP interface. Python version used for running the adaptation manager was 3.10.12.
The adaptation manager runs the $MSE$ loop every 200 seconds in our case. In this case, the objective given to the LLM is to keep the response time as low as possible.

\begin{figure}
    \centering
    \fbox{\includegraphics[width=3.2in]{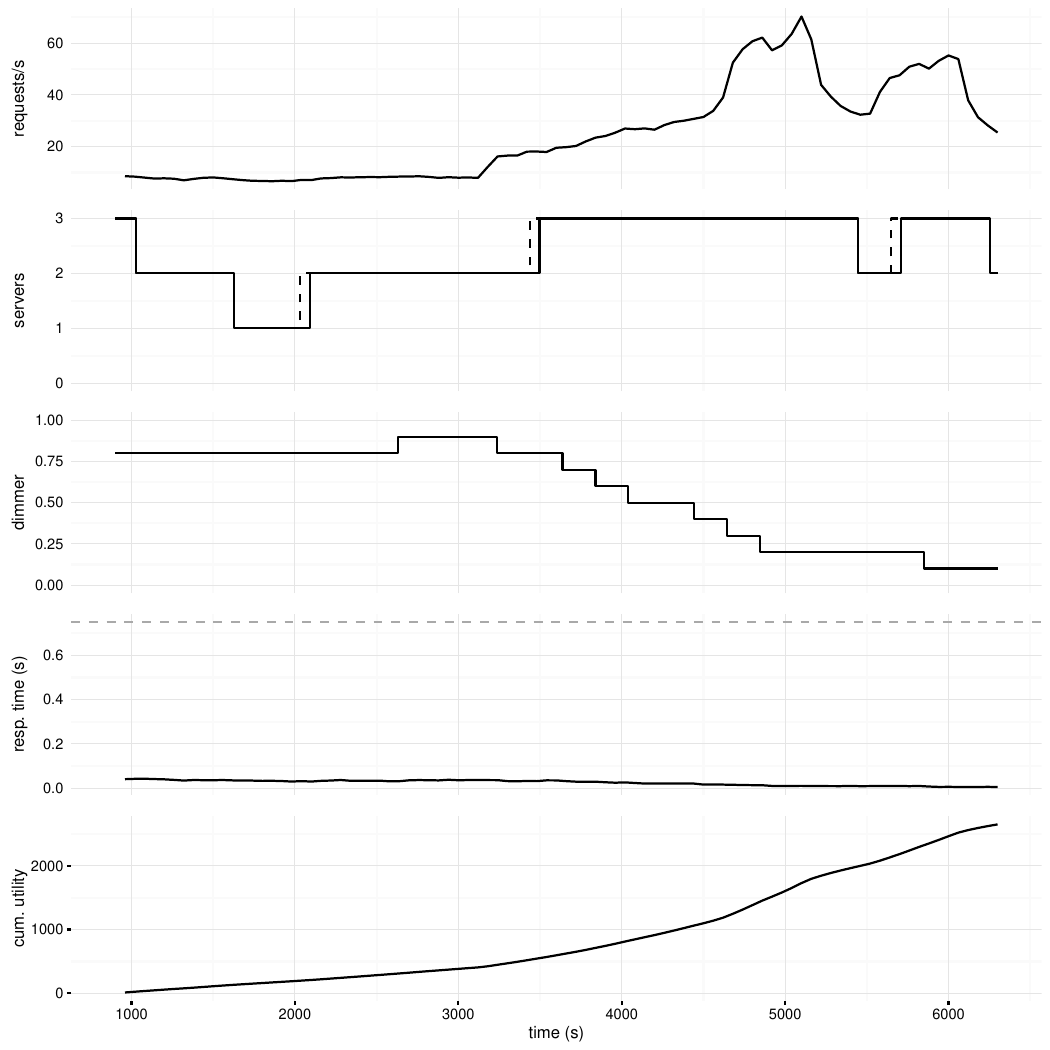}}
    \caption{Results from running the GPT-4 based adaptation manager on the world cup trace in SWIM. We observe that the average response time remains stable throughout the simulation.}
    \label{fig:Results}
\end{figure}
Our experiment \footnote[1]{\url{https://github.com/Raghav010/llm_selfAdapt}} 
involving SWIM demonstrates the potential of LLMs to enhance self-adaptation in software systems. As illustrated in Figure~\ref{fig:Results}, employing LLMs ensured stable response times below 0.1 seconds, even with varying workloads, achieving a utility score that exceeds a value of $2500$. When compared to the traditional reactive adaptation managers in SWIM~\cite{SWIM}, which although reaching a utility score exceeding a value of $3500$  struggled to maintain stable response times with spikes of more than $6$ seconds. The LLM-based adaptation manager attained a utility score that is roughly 71\% of the reactive adaptation manager. An increase in requests triggered the adaptation manager to efficiently scale server resources and decrease the dimmer value, thus minimizing the response time. These results effectively address our primary objective of reducing response time, highlighting the potential of LLMs to optimize resource allocation and system performance in real-time, as well as manage system behaviour dynamically.

\section{Vision for the Future and Further Research}
\label{sec:future_vision}
Exploring LLMs in self-adaptation marks the beginning of a promising research area with vast potential. Future efforts should focus on enhancing LLMs' understanding of complex system dynamics. While LLMs effectively grasp the qualitative effects of adaptations on system performance, their grasp of complex equations detailing adaptation impacts is limited. Integrating knowledge infusion techniques could provide LLMs with deeper insights into system operations and effects. Expanding to multiple LLMs or multi-agent LLM architectures~\cite{autogen} offers scalability and improved decision-making for comprehensive software applications, potentially reducing errors and improving adaptation quality. This setup could distribute system components across LLM agents or organize them hierarchically for deeper analysis.

Addressing long context challenges in production environments might benefit from advanced technologies like MemGPT~\footnote{https://github.com/cpacker/MemGPT} or StreamingLLM ~\cite{streamingllm}, focusing on optimizing scenario storage for better adaptation strategies. Fine-tuning LLMs, particularly for domain-specific applications, and employing Reinforcement Learning using System Feedback (RLSF) to adjust model parameters based on the system objectives, represent critical areas for future research. This approach could significantly enhance LLM performance and reliability.

Another challenge is the ever-discussed issue of hallucination which may result in incorrect decisions produced by the LLM. This is where we believe that LLMs combined with formal verification techniques similar to the approach presented by Camara et al.~\cite{9101287} can improve the guarantees on the decisions generated. 

Despite the list of aforementioned challenges, this paper underscores the transformative impact of integrating LLMs into self-adaptive systems, paving the way for more resilient, intelligent, and efficient software solutions. 

\section{Conclusion}
\label{sec:conclusion}
In this work, we have presented a novel self-adaptation approach that makes use of an MSE-K loop for dynamic architectural adaptation by leveraging the capabilities of LLMs.
Our initial findings through evaluations on an exemplar system highlight LLMs' transformative impact on software engineering, enabling complex, human-like decision-making and strategies for software to autonomously adapt their architecture with reduced human intervention. 
This research sets a foundation for further exploration into LLMs' capabilities, striving for software that is increasingly adaptive, resilient, and efficient in an ever-evolving technological landscape.


\bibliographystyle{IEEEtran}
\bibliography{main}




\end{document}